\documentclass[twocolumn,trackchanges]{aastex631}
\usepackage{amsmath}

\begin{document}

\title{Counting mass with Gaia: Mass Density of stars and stellar remnants in the solar neighborhood}

\author[0000-0002-5461-5778]{Artem Lutsenko}
\affiliation{Dipartimento di Fisica e Astronomia, Universitá di Padova, Vicolo Osservatorio 3, I-35122 Padova, Italy}
\affiliation{INAF - Padova Observatory, Vicolo dell’Osservatorio 5, I-35122 Padova, Italy}

\author[0000-0002-0155-9434]{Giovanni Carraro}
\affiliation{Dipartimento di Fisica e Astronomia, Universitá di Padova, Vicolo Osservatorio 3, I-35122 Padova, Italy}


\author[0000-0003-0250-6905]{Vladimir Korchagin}
\affiliation{Southern Federal University, Stachki 194, Rostov-on-Don 344090, Russia}

\author[0000-0003-2695-864X]{Roman Tkachenko}
\affiliation{Southern Federal University, Stachki 194, Rostov-on-Don 344090, Russia}

\author[0000-0001-5598-8720]{Katherine Vieira}
\affiliation{Instituto de Astronom\'ia y Ciencias Planetarias, Universidad de Atacama, Copayapu 485, Copiap\'o 1531772, Chile}



\begin{abstract}
In the light of new full-sky surveys, many attempts of creating a consistent Galactic model were made. The main interest is to estimate the still poorly understood dark matter content. However, the results vary depending on methodology, assumptions, and baryonic distribution used.
To understand this discrepancy, we take the first step by estimating the model-free local mass density of stars and stellar remnants. We use a complete sample from the Gaia Catalogue of Nearby Stars within 100 pc from the Sun, together with the data on the sample of the White Dwarfs and the recent estimate of the Neutron Stars concentration in the solar neighborhood. After correction for unresolved binary stars and accounting for missing low-mass stars, we find the local mass density of stars and stellar remnants
in the solar neighborhood is $\rho_{100} = 0.040^{+0.012}
_{ -0.006}M_\odot pc^{-3}$ with Kroupa IMF, and $\rho_{100} = 0.037^{+0.012}_{ -0.006}M_\odot pc^{-3}$ with Chabrier IMF. 

\end{abstract}



\section{Introduction} \label{sec:intro}
One of the key goals in studying the Milky Way galaxy is understanding how baryonic and dark matter are distributed in its subsystems. This knowledge 
is important not only for comprehending the Milky Way dynamics and its origin but also for improving our understanding of dark matter's nature. 

The approaches to estimate the dark matter's mass in the Milky Way galaxy are based on tracing the difference between the mass derived from the rotation curve or the assumption of equilibrium of the Galactic disk in the direction perpendicular to its plane and an estimate of the role of baryonic matter in establishing the equilibrium of the Galactic disk. There is a degeneracy between the estimate of the dark-matter content and the content of baryonic mass in the solar neighborhood. To get a reliable estimate for the dark-matter content, one should restrict the uncertainty in the density and the distribution of baryonic matter in the Galaxy by other methods.

An essential input in solving this problem is the knowledge of the mass density of stars, stellar remnants, and gas in the Galactic disk.
Information about baryonic matter in the solar neighborhood is required, for example, to infer the amount of dark matter in the solar vicinity. This can be achieved by modeling the distribution and kinematics of observational quantities for stars and gas in a combined gravitational potential of the dark matter and the baryonic matter. The model can be local, based on fitting the observational data for the stars in the solar vicinity, or based on fitting the rotation curve of the Galaxy for a large range of distances. 
Knowledge of the density and concentration of the dark matter in the solar neighborhood is important for estimating the sensitivity requirements in the experiments for dark matter searching and for constraining the shape of the Milky Way dark matter halo.

Studies of the stellar mass budget in the solar neighborhood have a long history starting from the work of Jakob Kapteyn \citep{Kapteyn22}.  Useful references on studying the budget of stars and the initial mass function in the pre-Gaia era can be found in the review paper \citet{Read14}. A comprehensive study of the stellar mass density in the solar neighborhood in the pre-Gaia era was performed by \citet{Kee15}. These authors found that the local mass density of stars in the solar neighborhood is $0.043M_\odot pc^{-3}$.

The launch of the Gaia satellite triggered a surge of interest in the study of stellar populations and the initial mass function. After the first preliminary data release of Gaia data, \citet{Bovy17} estimated the midplain stellar density in the solar neighborhood to be $0.040 \pm 0.002 M_\odot/pc^3$. \citet{Widmark} used data from the Second Data Release of the Gaia mission and determined the density of gravitating matter in the solar neighborhood.  \citet{Widmark} estimated local mass density and found that there is a surplus of midplane mass density of approximately $0.05-0.15 M_\odot/ pc^{-3}$, compared to the baryonic density budget taken from \cite{Schutz}, who in turn compiled data for different baryonic components of \citet{Flynn}, \citet{Kee15}, and \citet{Kramer}. \citet{deSalas2019JCAP...10..037D} compared two different models for the baryonic mass distribution and found that the estimation of the dark matter density is sensitive to the assumed baryonic distribution with the one sigma uncertainty of the local dark matter density reaching $0.004M_\odot pc^{-3}$, or up to $50\%$ of the local density of dark matter.

{\citet{Widmark2021A&A...646A..67W} inferred the Galactic gravitational potential from the velocity distribution of stars extracted from Gaia Date Release 2 (DR2) assuming steady state. They used this potential to derive the baryonic mass profile of the disk. They obtained a steep density profile, with a lack of material at large distances from the plane and a strong concentration of mass very close to the plane. Instead of attributing these results to a lack of mass away from the plane, they suggested that this profile reflects local time-varying dynamical perturbations.

To test how assumptions affect an estimate of dark matter density, \citet{Sivertsson2022_10.1093/mnras/stac094} performed an N-body simulation of the dynamics of the Milky Way-like galaxies that experienced a major merger of the  Gaia-Sausage–Enceladus type or a recent merger with a dwarf galaxy similar to Sagittarius. The authors concluded that commonly used assumptions in the literature, such as an axisymmetry of the disk and the locally flat rotation curve, lead to significant systematic errors in the mass density estimate, while steady-state approximation still allows us to derive the correct estimate of the dark-matter density within the $95\%$ confidence interval.

The third preliminary release of the Gaia catalog has enabled a more detailed study of the stellar content and the stellar distribution function in the solar neighborhood. 

Using standard quality cuts and taking into account the selection function, \citet{Everall2022} fitted the parameters of the model consisting of exponential thin and thick disks as well as a power-law halo. The authors analyzed several systematic effects and concluded that systematic errors dominate statistical ones. They find that the stellar mass density excluding compact objects is $0.0366\pm0.0052M_\odot pc^{-3}$.

\citet{Kirkpatrick_2024} (hereafter K24) analyzed Gaia Data Release 3 (DR3) data together with ground-based surveys and built a volume-limited sample of $\approx 3600$ individual stars within a sphere of 20 pc radius centered at the Sun. The sample allowed the authors to reevaluate the stellar mass distribution down to the late L-type stars. For colder stars, K24 used data from the Wide-field Infrared Survey Explorer (WISE) together with parallaxes measured by Spitzer. To build a sample of stellar objects within 20 pc of the Sun, K24 split the binary and multiple systems into their individual components. However, the authors noticed that there are likely undiscovered brown dwarfs that are too faint to be detected currently. This can affect the final form of the present-day stellar mass function in the solar neighborhood.

The volume-limited sample used by K24 includes only a few stars with masses larger than $2 M_\odot$. The number of stars with lower masses can be affected by the local fluctuations in the limited volume sample selected by K24. 

We used the complete sample of 302449 stars from the Gaia Catalog of Nearby Stars (GCNS) \citep{Smart_GCNS_2021} taken within the sphere of 100 pc radius centered at the Sun. The density of white dwarfs in the solar neighborhood was estimated using recent data from \citet{O’Brien_WD40_10.1093/mnras/stad3773}. To estimate the contribution of neutron stars, we use the local surface density of \citet{Xie_2024_NS}, aiming to provide the observationally based value of the local stellar density, which is independent of the Galactic disk model. 

The paper is presented in the following way. Section \ref{sec:data} describes the observational data; in Sect. \ref{sec:mass} we explain how the masses of the stars are calculated. Section \ref{sec:Corrections} describes the corrections that should be made to obtain the mass density in the solar neighborhood. The results of our study are presented in Sect. \ref{sec:res}. We compare our results with previous works in Sect. \ref{sec:disc}. Section \ref{sec:sum} summarizes our findings.

\section{Data selection} \label{sec:data}

We choose for our analysis the Gaia Catalog of Nearby Stars (GCNS) \citep{Smart_GCNS_2021}, which is based on the Gaia Early Data Release 3 \citep{Gaia_EDR3}. We strive for a maximum completeness level of a wide range of magnitudes, and hence we have to select the stars in the solar vicinity. Some sources might have large parallaxes due to the imprecise astrometric solution and can be identified mistakenly as nearby stars. The general recommendation to eliminate such sources is to use Renormalized Unit Weight Error (\textit{ruwe}) as a quality indicator of the astrometric solution. However, this quality cut leads to the loss of binary stars that truly reside in the solar vicinity.\citep{Penoyre2022}. 
Some sources were processed by the pipeline erroneously, so they do not represent the stars and should be excluded from the analysis.

\citet{Smart_GCNS_2021} provided a table that was already cleaned by the random forest classifier from spurious sources, as well as distant ones that have large parallaxes due to poor astrometric solutions.
Avoiding the usual quality cut using the value of \textit{ruwe}, \citet{Smart_GCNS_2021} managed to keep nearby sources with large values of \textit{ruwe}, likely being unresolved binary systems of stars. However, we should mention that some faint stars ($G\_abs > 14$) were likely removed because of the absence of 2MASS magnitudes (see Section 2.1 and Fig. 2 in the paper for more details).

\citet{Smart_GCNS_2021} estimated the completeness of their catalog about 98\% for the apparent magnitude $G=20.48$, which converts to the absolute magnitude of $G_{abs}=15.48$ at a distance of 100 pc. Thus, the GCNS is not magnitude limited for stars brighter than this threshold. After the random forest classifier \citet{Smart_GCNS_2021} estimated the contamination of their catalog of about $0.1\%$ due to the false positives of the random forest classifier, and about $9\%$ due to the stars outside of the 100pc sphere. The latest will be removed later in the work.

Thus, the GCNS provides a good balance between purity and completeness for a wide range of absolute magnitudes, which is very convenient for our purpose.

The GCNS includes a distance estimate determined with the help of the Bayesian method. To reconstruct the absolute magnitudes, we use the median value of the posterior distance estimate $dist\_50$, also assuming the absence of extinction.

The white dwarf (WD) population of stars in the GCNS requires an independent investigation, so in this work we rely on results obtained by \citet{O’Brien_WD40_10.1093/mnras/stad3773}, who studied the White Dwarf sample of stars within 40 pc around the Sun, extrapolating the obtained WD population mass density to the sphere of 100 pc radius around the Sun (see details in Sect. \ref{subsec:WD_NS_BH}). For the estimation of the density of neutron stars, we use the surface density value obtained by \citet{Xie_2024_NS} for pulsars after beaming correction.

Excluding about $2\%$ of stars from the GCNS catalog that do not have photometric data, and removing about $7\%$ of the white dwarfs by the cut shown in Fig. \ref{fig:cmd_cut}:
\begin{equation}
RP_{abs} - 8 \cdot (G - RP) - 6 < 0
\end{equation}
where the $RP_{abs}$ is an absolute RP magnitude determined with help of $dist\_50$ - median value of the posterior distance estimate:
\begin{equation}
RP_{abs} = RP - 5 \cdot lg(1000 \cdot dist\_50) + 5
\end{equation}
After the cut, the sample of the stars consists of 302449 sources. Additionally, we cross-matched our sample with \citep{stellar_par} to obtain the stellar parameters, which we use to derive masses in Sect. \ref{subsec:mass_stellar_par}.

\begin{figure*}
  \resizebox{\hsize}{!}{\includegraphics{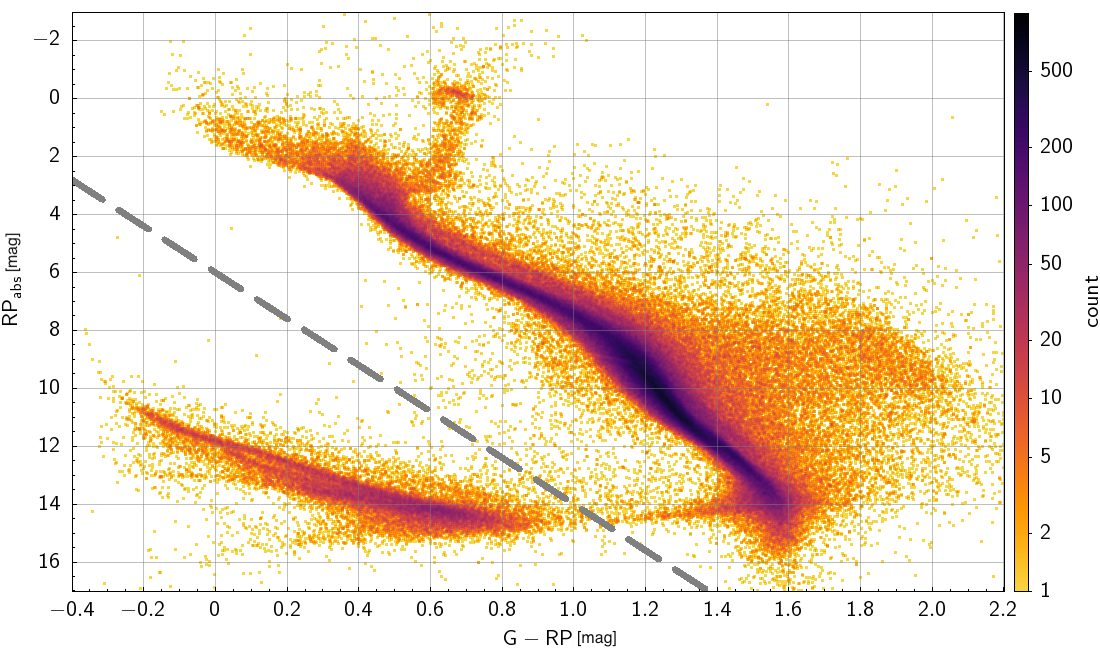}}
  \caption{Color absolute magnitude diagram of the GCNS for absolute magnitude $RP_{abs}$ and color index $G - RP$. The gray dashed line is the line following $RP_{abs} - 8 * (G - RP) - 6 = 0$.}
  \label{fig:cmd_cut}
\end{figure*}

\section{Masses of individual stars} \label{sec:mass}

An estimation of the mass of individual stars is usually done by employing evolutionary stellar tracks. Unfortunately, reliable isochrone fitting for the field stars is indeed a very complex task due to the complex star formation history of our Galaxy. Stars of different ages and different chemical abundances are mixed, making it difficult to differentiate the turn-off point of the main stellar population. In addition, stellar rotation and close binary stars contribute to blurring the main sequence and its turn-off point. In such a complex environment, an isochrone fitting technique is not precise, and several combinations of stellar tracks with different ages and metallicities can mimic the same distribution on the color-magnitude diagram.

Another way to obtain stellar masses is to use the empirical relations between magnitudes or other stellar parameters and masses reconstructed from the orbital motion in binary systems. The obvious constraint of this method is that a star should belong to a binary system in order to obtain the mass estimation. The fraction of binaries varies at different parts of the color-magnitude diagram, making it difficult to obtain a large data set to build a mass-magnitude relation for a wide range of magnitudes.

In this study, to circumvent the complexity of isochrone fitting, we derive masses for individual stars with absolute magnitudes $RP_{abs}$ in the Gaia red photometric filter larger than 4 magnitudes employing an empirical mass-magnitude relation. For brighter stars, we used the relation between mass, effective temperature, surface gravity, and metallicity described in the following subsection.

\subsection{Mass-Magnitude relation} \label{subsec:Mass-Magnitude relation}

To derive the mass of a star from its absolute magnitude, we use the mass-magnitude relation taken from \citet{Giovinazzi_2022} (hereafter G22). The relation is based on the analysis of more than 3800 wide binary systems. G22 reconstructed the orbital motion of the components in the binary systems and approximated the relation with an analytical formula. The G22 relation is valid in the red absolute Gaia magnitude $RP_{abs}$ ranging from 4.0 to 14.5.

For comparison, we also used a polynomial function taken from K24. This function is adopted for Gaia photometry from \citet{Mann_2019} and is derived using 62 nearby binary systems with well-determined orbital parameters. 
The applicable range of absolute Gaia magnitudes $G_{abs}$ goes from 7.5 to 15.0.


The absolute magnitude for a source was derived by using the apparent magnitude and a posterior distance estimation. 
The apparent magnitude value was generated from the Gaussian distribution using the value of apparent magnitude reported in the catalog as the mean and the reported error as a standard deviation. Equation \ref{eq:gen_rp} represents the probability distribution function used for the procedure to generate the magnitude $RP$ in Gaia notation:

\begin{equation}
\begin{split}
RP =  &Norm(phot\_rp\_mean\_mag, \\
& \frac{2.5}{ln(10) \cdot phot\_rp\_mean\_flux\_over\_error} )
\end{split}
\label{eq:gen_rp}
\end{equation}

Similarly, we generate distance using a Gaussian distribution with the mean posterior distance estimation as a mean and the maximum posterior distance estimation at 16 and 84 percentiles as a standard deviation. We did not use an extinction correction in this study.

Taking the stars from the common range of the magnitudes for both relations, we derive the distribution of mass that generates apparent magnitudes and distances $10^3$ times for each star. The individual mass of a star is derived as the mean value of the distribution and the error is derived as the standard deviation (std). Illustrative examples are shown in Appendix \ref{sec:ind_mass}. 

The mass-magnitude relations in their common validity range from 7.5 to 15 absolute magnitudes $G_{abs}$ ($G_{abs} \in [7.5,15]$) give similar results. We find that the total mass difference between the mass-magnitude relations is about 1\%. $(67.4\pm 9.6)\times10^3 M_{\odot}$ is the total mass using the G22 relation and $(66.6\pm 6.2)\times10^3 M_{\odot}$ is the total mass using the K24 relation (see Fig.\ref{fig:mass-mag} left panel). 


Despite the smaller errors of the K24 relation, in this study we use the mass-magnitude relation of G22 because of its large absolute Gaia magnitude $RP_{abs}$ range from 4.0 to 14.5.

\begin{figure*}
  \resizebox{\hsize}{!}{\includegraphics{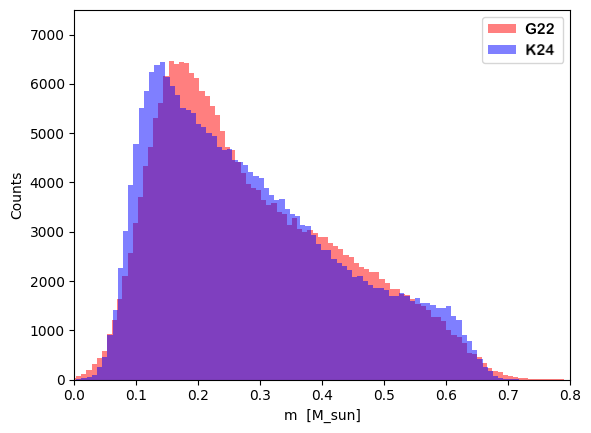}
    \includegraphics{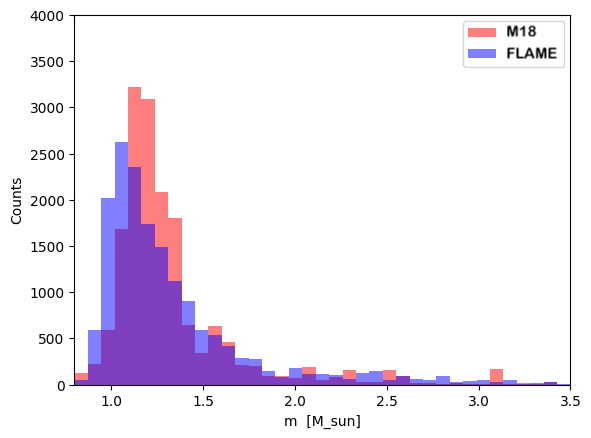}}
  \caption{Histogram of the mass of stars. Left panel: the stars in the absolute magnitude $G_{abs}$ range from 7.5 to 15. Red is derived from the G22 mass-magnitude relation, blue is derived from the K24 mass-magnitude relation. Right panel: the stars in the absolute magnitude $RP_{abs}$ lower than 4. Red is derived from the M18 relation between mass and stellar parameters, and blue is derived using the FLAME masses. The histograms represent only one observable generation.}
  \label{fig:mass-mag}
\end{figure*}

\subsection{Mass from astrophysical parameters} \label{subsec:mass_stellar_par}

 During its evolution, a star experiences a significant change in luminosity and mass when it departs from the main sequence. The mass-magnitude relation gives only a rough estimate of the mass of the evolved stars. We therefore use the relation between mass and stellar parameters, such as effective temperature, surface gravity, and metallicity. 

\citet{Moya_2018} (hereafter M18) provided a set of relations between stellar parameters and mass. These authors used 934 MS/post-MS stars with precise stellar parameters derived from astero-seismology data, eclipsing binaries, and interferometry. After regression, M18 derived the dependence of a stellar mass as a function of different stellar parameters. We use the relation $log(m) \sim T_{eff} + log(g) + [Fe/H]$  because it has one of the highest relative precisions and relative accuracy (see Table 5 of M18).

We do not have stellar parameters for all stars with absolute Gaia magnitude $RP_{abs}$ smaller than 4. However, this region of the color-magnitude diagram is dominated by bright stars and, therefore, we have stellar parameters for most stars ($\approx 87\%$).

There are two sets of stellar parameters provided by \citet{stellar_par}. The first set is called $gsp\_phot$ and is derived from the $BP$ and $RP$ spectra, while the second set is $gsp\_spec$ and is derived from the $RVS$ spectra. The second set is more precise but is available for a smaller number of stars, about $\approx 65\%$ compared to $\approx 87\%$ for the first set of data.

In order to compute the mass of all stars in this region, we create a grid binned with a $0.2$ magnitude step in color index $BP-RP$ and a $0.5$ magnitude step in absolute magnitude $RP_{abs}$ (see Fig. \ref{fig:cmd_above4}). We generate the observable values using the recipe explained in Sect. \ref{subsec:Mass-Magnitude relation} and distribute the stars on the grid. To compute the masses of the stars, we use the following prescription.


\begin{figure*}
  \resizebox{\hsize}{!}{\includegraphics{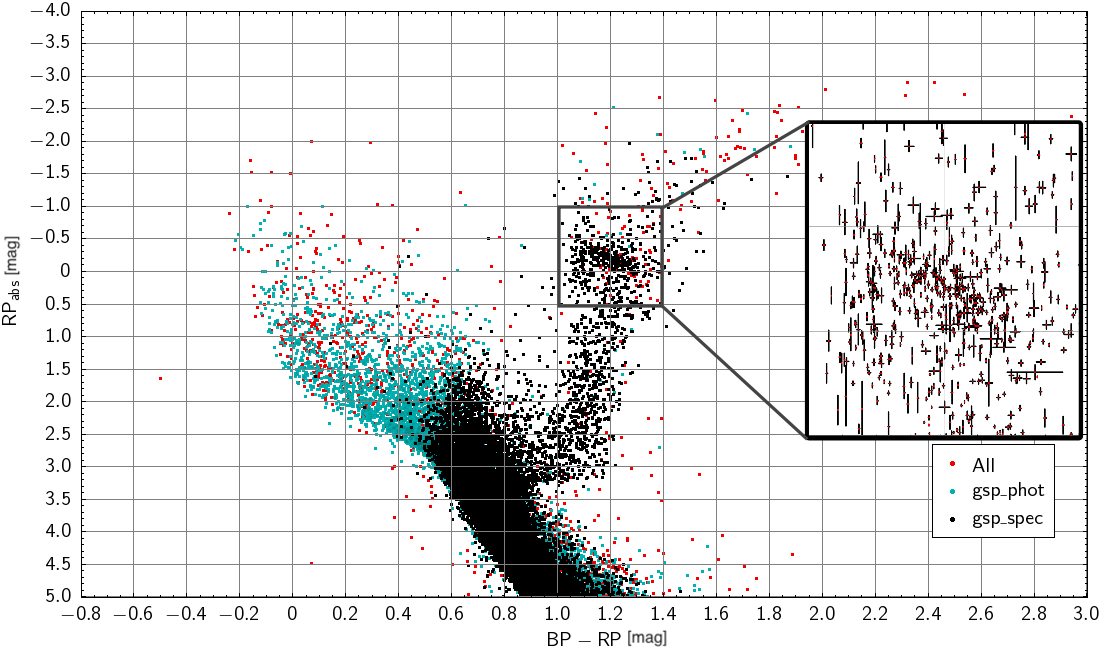}}
  \caption{Color absolute magnitude diagram with the net used for the calculations. Red dots are all the stars, cyan dots are the stars with $gsp\_phot$ stellar parameters, and black dots are the stars with $gsp\_spec$ stellar parameters. Inside the zoomed-in window, one can see error bars propagated from errors on distance and apparent magnitudes.}
  \label{fig:cmd_above4}
\end{figure*}

\begin{enumerate}
  \item We computed the mass using the M18 relation with the stellar parameters $gsp\_spec$ ($\approx 65\%$).
  \item For stars that do not have $gsp\_spec$, we compute the mass using the M18 relation with the $gsp\_phot$ stellar parameters ($\approx 21)\%$.
  \item For stars that do not have stellar parameters, we assign the mass as the average mass inside the cell ($\approx 13\%$).
  \item For cells that do not have stars with stellar parameters, we assign the average mass of the cell as the average mass of the neighbor cell in the same magnitude line ($<1\%$).
\end{enumerate}

We repeat the procedure $10^3$ times, generating the magnitudes, distance, and stellar parameters as explained in Sect. \ref{subsec:Mass-Magnitude relation}. Then the mass of individual stars is calculated as a mean of the mass distribution, while the error is calculated as std. Examples can be found in Appendix \ref{sec:ind_mass}.

In addition, we compared the mass obtained with our procedure with the \textit{Final Luminosity Age Mass Estimator} (FLAME) masses \citep{stellar_par}, which are available for $\approx 86\%$ stars. Deriving the mass for the stats without FLAME mass as an average mass of the cell of the grid, and again deriving errors with the bootstrapping method generating observables and FLAME masses $10^3$ times, we confirm that the total mass difference is less than $1\%$ (see Fig.\ref{fig:mass-mag} right panel), $(22.5\pm2.6)\times10^3 M_{\odot}$ using the M18 relation and $(22.6\pm 1.6)\times10^3 M_{\odot}$ using the FLAME masses. To assess the robustness of the binning scheme, we shifted the grid along both axes by 0.2, 0.4, 0.6, and 0.8 of the bin width, resulting in 25 different configurations. Using the difference in the shifted configurations with respect to the initial one, we confirm that the maximum absolute error related to the binning is $55.6M_{\odot}$. Therefore, the final error is dominated by errors in observational values and in the relation between mass and stellar parameters. This is reasonable considering that the binning affects only 14\% of stars. We should mention that for the mass computation, we used all available FLAME masses, regardless of the quality flag or the mass being larger than $2M_{\odot}$ (for more details, the reader can consult \citet{stellar_par} Sect. 6.4.1). In this work, we use only the M18 relation. Examples of individual mass estimated using the M18 relation and using the FLAME masses are shown in the Appendix \ref{sec:ind_mass}. Perhaps, the mass derivation could be imprecise for stars without stellar parameters; however, we highlight that we are interested in the total mass estimation.

\section{Stellar mass function corrections} \label{sec:Corrections}

As was mentioned above, the GCNS catalog is not complete for all the range of absolute magnitudes. We should therefore determine an effective volume in which the GCNS catalog is magnitude complete for a specific type of star. The low-mass faint stars that have absolute magnitudes $G>17$ can be observed only at small distances. K24 used in their study the 20 pc sphere centered at the Sun. However, even this cannot prevent the incompleteness of faint-end stars in the crowded areas of the sky. Also, the close binary systems not resolved into individual stars lead to the loss of mass of the second component. All of these factors can cause a significant underestimate of the mass density of the stars in the solar neighborhood and should be taken into account.

\subsection{Volume correction}
\label{subsec:vol}

Due to the apparent magnitude limits of Gaia, we have stars within the range of apparent magnitudes $3 \le G \le 21$. The distance modulus at 100 pc is equal to 5 mag, hence, the volume-complete range of absolute magnitudes of the stars should be from 8 to 16. In Figure \ref{fig:vol_cor}, one can see how the stars with absolute magnitude $G_{abs}$ are distributed with distance. The black lines correspond to the apparent magnitude limits, while the lines highlighted in blue are Gaia limits with 3 and 21 magnitude limits. Stars that have an absolute magnitude $G_{abs}$ lower than 8 are too bright to be observed by Gaia close to the Sun, which justifies the absence of stars above the blue line from the top (see the top of Fig. \ref{fig:vol_cor}). Stars that have absolute magnitudes $G_{abs}$ greater than 16 are too dim to be observed at large distances, which is evident from the absence of stars below the bottom blue line (see the bottom of Fig. \ref{fig:vol_cor}).

\begin{figure}
  \resizebox{\hsize}{!}{\includegraphics{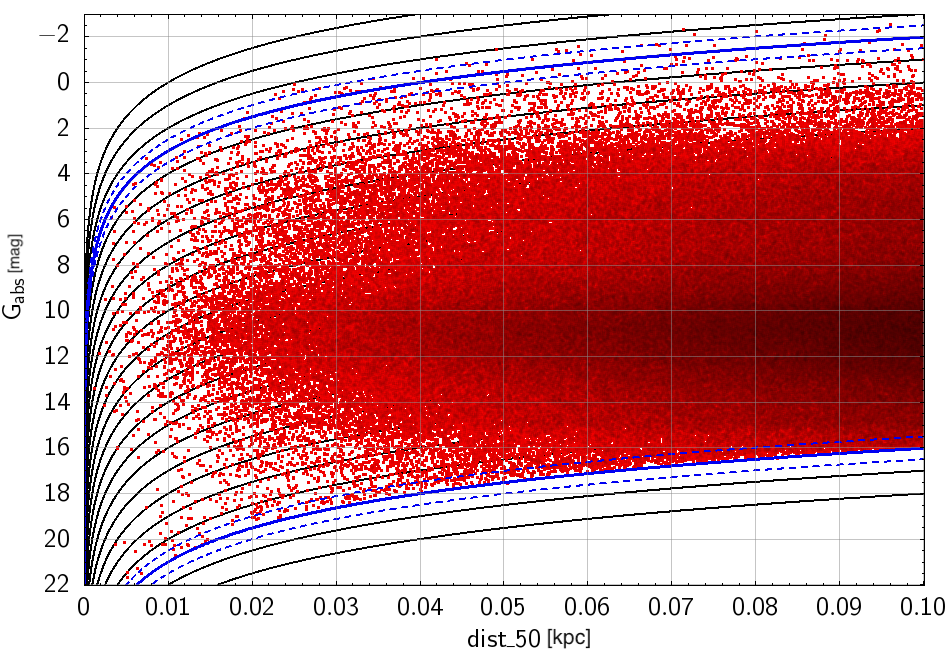}}
  \caption{Absolute magnitude $G_{abs}$ vs distance $dist\_50$. The red points are the stars from GCNS sample excluding WD stars. The black lines are distances derived from different apparent magnitude $G$ limits according to Eq. \ref{eq:R}, the lines highlighted blue correspond to the limits of 3 and 21, with half of the magnitude error plotted by blue dashed lines.}
  \label{fig:vol_cor}
\end{figure}

To select the volume-complete samples of stars with different absolute magnitudes, we used the following procedure. For stars that have absolute magnitudes $G_{abs}>15.5$, we use an effective spherical volume determined by the radius $R$

\begin{equation}
\label{eq:R}
R = 10^{\frac{G_{lim}-G_{abs}}{5}+1},
\end{equation}

where $G_{lim}=21\pm0.5$  is the upper apparent magnitude limit. The errors are used to derive the lower and upper radius, which are propagated in the calculation of the effective volume and, therefore, density.


Stars with absolute magnitudes $G_{abs}<7.5$ are selected in the volume determined as:

  \begin{equation}
  {V_{eff}} = \frac{4\pi}{3}\cdot (100^3 - R^3), 
  \end{equation}
  
where the radius $R$ is determined by Eq. \ref{eq:R} and $G_{lim}=3\pm0.5$ is the lower apparent magnitude limit. 

Figure \ref{fig:vol_cor_2} presents the normalized cumulative histograms for three sub-samples of stars with $15.9 < G_{abs} < 16.1$, $16.9 < G_{abs} <17.1$, and $17.9 < G_{abs} < 18.1$ correspondingly. As one can see, at the distance obtained from  Eq. \ref{eq:R}, the cumulative distribution reaches its maximum, which means that there are no stars after this distance. In addition, the histograms follow the distance-cube line, which means that the sub-samples are complete. However, for the fainter subsample, we see the most significant deviation from the distance-cube line due to the smaller total number of stars in the subset.



    

\begin{figure}
  \resizebox{\hsize}{!}{\includegraphics{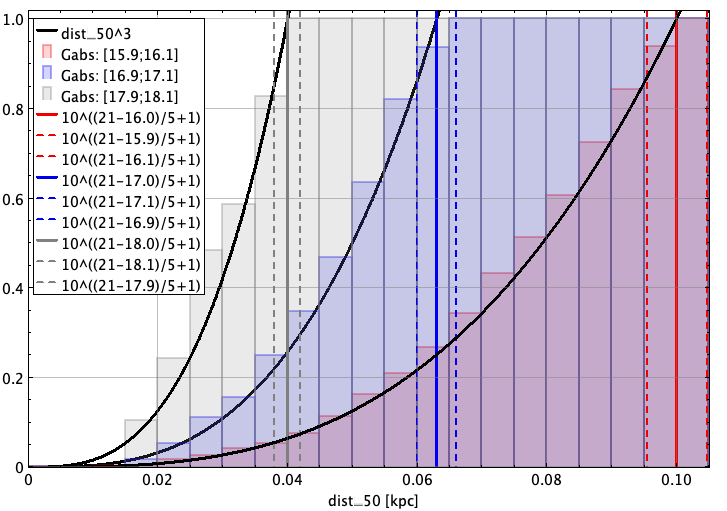}}
  \caption{Normalized cumulative histograms of the subsamples. The bin size is 5 pc. The red histogram is a subsample with absolute magnitude $G_{abs}$ in the range $[15.9;16.1]$, while the red vertical lines correspond to the distance derived from Eq. \ref{eq:R} for the absolute magnitudes in the subsample. Likewise, the other two subsamples and their distances are marked. The black lines are proportional to the distance cube and normalized at the distance corresponding to each subsample. The total number of subsamples is 977, 173, and 29, respectively.}
  \label{fig:vol_cor_2}
\end{figure}

\subsection{Binary correction}
\label{subsec:binary}

The correction takes into account a missing mass due to unresolved binary systems. Recently, it was shown that the value of the high renormalized unit weight error \textit{ruwe} is a good tracer of a binary system \citep{Penoyre2022,Belokurov2020}. 
Since the \textit{ruwe} represents the quality of the astrometric solution, for areas of the sky with fewer observations, the \textit{ruwe} value of single stars occupies a larger interval, and thus the threshold after which we expect a star to be binary is higher. For the crowded regions as per say the direction of the Galactic center, the light of a star can be contaminated by the unresolved background, so for such regions the \textit{ruwe} of single stars occupies a wider interval, and, as a consequence, the threshold is also higher. 

We used the sky-dependent \textit{ruwe} threshold provided by the GaiaUnlimited Python package \citep{Castro-Ginard_ruwe}, where the authors took care of the Gaia scanning law and crowded regions to derive the threshold.
We assume that a star is a close binary system if its \textit{ruwe} value is larger than the threshold for a particular location on the sky. Using this criterion, we find that $\approx 24\%$ of the stars are close binaries. The value is consistent with a recent estimate for the solar neighborhood \citep{Wallace2023}.
The binaries occupy the entire range of absolute magnitudes, but the most prominent region is MS shifted up, where we expect binary systems to be (see Fig. \ref{fig:binary}).
Sources with a color index $G-RP$ in the area of $1.8-2.0$ mag (see Fig. \ref{fig:binary}) have a large color excess and $ipd\_frac\_multi\_peak$ more than 50\%, also indicating that they are likely binaries \citep{Clark_2024}.

The mass of the second component in a binary system depends on the mass of the primary star.  
However, the dependence is still a subject of discussion whether the distribution of mass ratio is flat or power-law or whether it depends on the orbital period or the type of the primary star \citep{Duchene_bin_2013ARA&A..51..269D,Reggiani_bin_2013A&A...553A.124R}. Another discussion is whether we can apply the same relations measured for open clusters to the field stars. Since the solar neighborhood consists not only of young stars but also of old stars, the binary fraction and distribution of the mass ratio are likely to be different due to the decay of binary systems with time \citep{Korntreff2012,Donada_bin2023}.

We assume here that the mass of the second component is equal to half the mass of a primary star. The more accurate estimation requires deep investigation between \textit{ruwe} and properties of binary systems, such as period, orbit inclination, and mass ratio.


\begin{figure}
  \resizebox{\hsize}{!}{\includegraphics{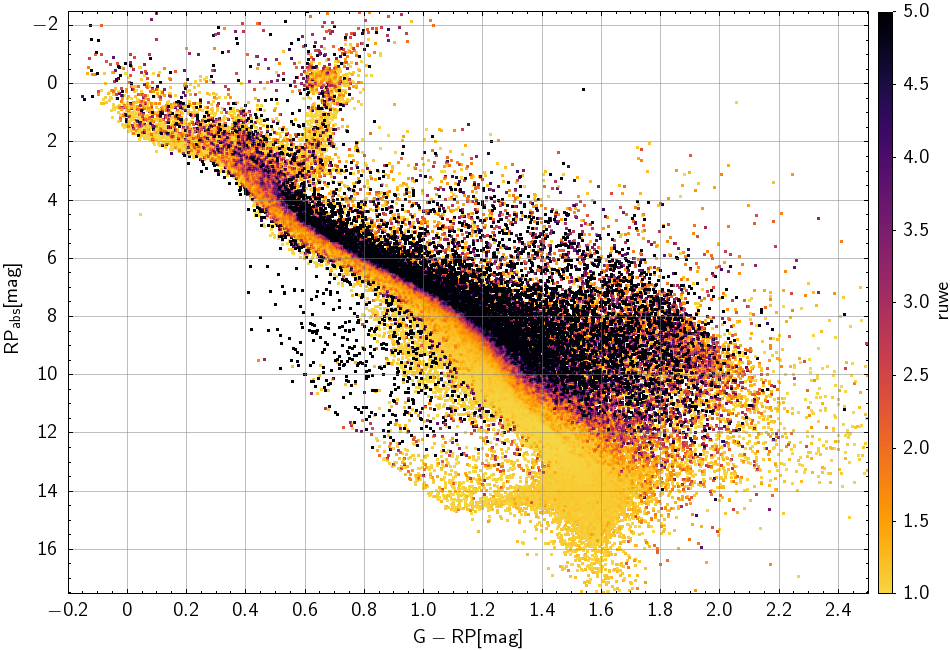}}
  \caption{Color absolute magnitude diagram color coded by \textit{ruwe}.}
  \label{fig:binary}
\end{figure}

\subsection{Low mass correction}
\label{subsec:imf}

After taking into account volume correction, we still lose completeness in the sample of stars at the low-mass end of the main sequence because of selection effects and sky crowding. According to the Gaia limiting magnitude, the effective volume of faint stars is small; therefore, the number of stars is also small. So, losing even a few stars due to the crowding of false classification by the random forest classifier in \citep{Smart_GCNS_2021} can severely affect the mass function and therefore the mass density estimation.

To correct the mass function for the low-mass end of the main sequence, we use the Initial Mass Functions (IMF) of \citet{Kroupa2001MNRAS.322..231K} (hereafter K01) and \citet{chabrier2005initial} (hereafter C05), taking into account that for the long-living low-mass stars the IMF is similar to the present-day mass function. 
To determine the stellar distribution, we use the maximum likelihood estimation technique, with the amplitude of the IMF being the only free parameter. Fixing the shape of the IMF and altering only its amplitude, we find that the amplitudes are $k=0.0140\pm0.0001$ in the case of the K01 IMF and $A=0.093\pm0.006$ for the C05 IMF. The resulting distribution of stellar mass for both initial mass functions is shown in Fig. \ref{fig:mass_hist}.

As can be seen in Fig. \ref{fig:mass_hist}, both functions agree well with our data for stars with mass $m > 0.2 M_{\odot}$ and differ considerably from the observational data for low-mass stars because of the selection effects.


\begin{figure*}
  \resizebox{\hsize}{!}{\includegraphics{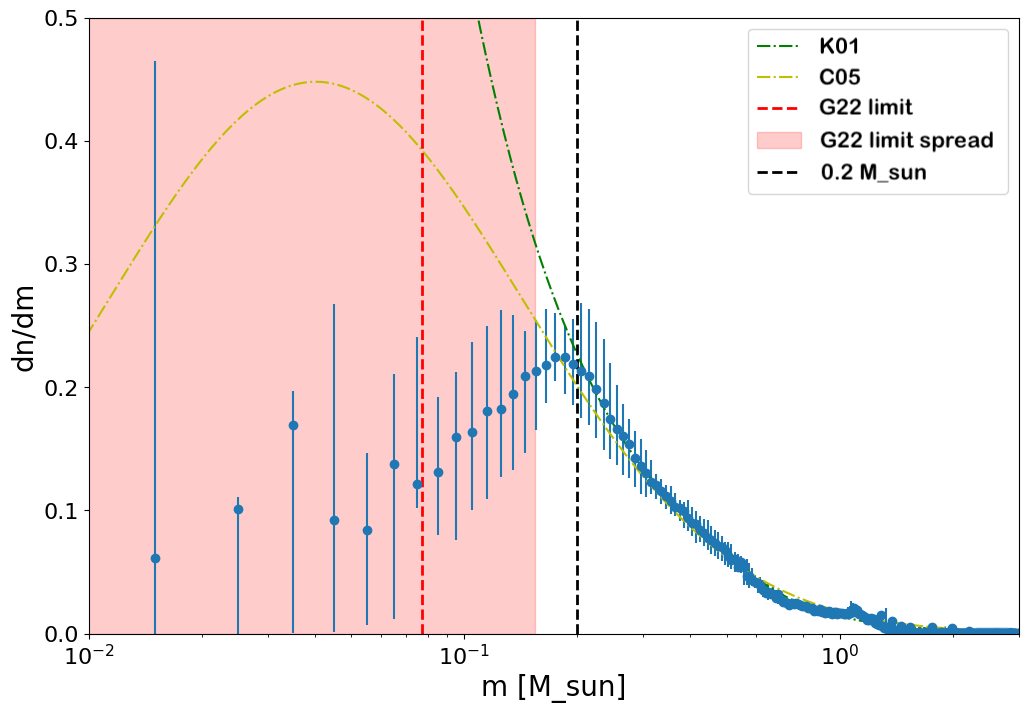}}
  \caption{Distribution of the mass derived for $0.01M_\odot$  bin size corrected for volume-incompleteness and binaries. The blue points are the data derived from this work with error bars propagated from errors of mass and effective volume derivation. The green line is K01 IMF. The yellow line is C05 IMF. The red dashed line represents the limit of the mass-magnitude relation at $RP\_{abs} = 14.5$, and the red area is the spread of the limit according to the error of the relation. The black dashed line shows the threshold of $0.2M_\odot$.}
  \label{fig:mass_hist}
\end{figure*}

\section{Results} \label{sec:res}

\subsection{MS and evolved stars}

Here we estimate the mass density of stars with a mass bigger than $0.2 M_{\odot}$. Depending on the absolute magnitude $RP_{abs}$, the mass of the stars is computed using the mass-magnitude relation or the relation between the mass and stellar parameters. The mass of stars with $RP_{abs}>4$ is estimated as explained in Sect. \ref{subsec:Mass-Magnitude relation}, and for stars with $RP_{abs}<4$, the mass is derived in Sect. \ref{subsec:Mass-Magnitude relation}. Stars that lay outside of a 100pc sphere were removed.

Then we obtain the effective volume for each star as described in Sect. \ref{subsec:vol}, and build the mass function using the effective volume to derive the weight for a star as 

\begin{equation}
w = \frac{V}{V_{eff}}
\end{equation}

Therefore, we obtain the mass density as a sum of the weighted mass divided by the volume of the sample. 

\begin{equation}
\rho= \sum_i\frac{M_i}{V_{eff,i}} = \frac{1}{V}\sum_iM_i\cdot w_i
\end{equation}

As described in Sect. \ref{subsec:imf}, the drop after $0.2 M_{\odot}$ is due to selection effects, so we sum the masses larger than this threshold. The resulting mass density is $0.025_{-0.006}^{+0.012} M_{\odot}/pc^3$

\subsection{Binaries}
Following the prescription of Sect. \ref{subsec:binary}, we obtain the mass of the second unresolved component of binary systems. The contribution of the correction for masses larger than $0.2 M_{\odot}$ is $0.0034\pm0.0003 M_\odot/pc^{3}$.

\subsection{Faint end of MS}

The weighted histogram of the mass with 0.01 bin size is presented in Fig. \ref{fig:mass_hist}. The error bars correspond to the mass errors and the error of magnitude limits in the derivation of the effective volume and, therefore, the weight. The mass distribution function is already corrected for the mass of the second component of binary systems. To compute the mass density of the low-mass end of the mass distribution, we rely on the widely accepted IMF of K01 and C05 as explained in the Sect. \ref{subsec:imf}. Integrating the IMF multiplied by the mass from $0$ to $0.2M_\odot$ and dividing by the volume of our sample, we obtain the mass density of the low-mass end. 

With K01 IMF the mass density is $0.0089\pm0.0001 M_\odot/pc^{3}$, with C05 IMF the mass density is $0.0058\pm0.0001 M_\odot/pc^{3}$

\subsection{Stellar remnants} \label{subsec:WD_NS_BH}

To obtain the mass density of the WD population, we sum up the corrected masses from \citep{O’Brien_WD40_10.1093/mnras/stad3773}. For the 27 WD stars without mass estimation, we added their mass as the average mass of a WD of $0.6M_\odot$. Dividing by the volume of the sphere with a radius of 40 pc, we obtain the mass density of the WD population of $0.00269\pm 0.00009M_\odot/pc^{3}$

\citet{Xie_2024_NS} reported the surface density of pulsars around the Sun of $470\pm50 pulsars/kpc^2$. Conservatively assuming that all pulsars are concentrated in the thin disk with a vertical scale height of 0.28 kpc, we get $470 \pm 50 = N_0 \int 1/cosh^2 (\frac{x}{0.28}) dx$. Integrating this distribution between -2 and 2 kpc, we get the concentration of pulsars at the midplane of the disk (around the Sun): $N_0 = 840 \pm 89 kpc^{-3}$ which gives the mass density of pulsars around the Sun of $1176 \pm 124 M_\odot / kpc^3$ or $(1.176 \pm 0.124)\times10^{-6}M_\odot / pc^3$ assuming the average mass of $1.4M_\odot$. \citet{Arnaud1981} estimated the lifetime of pulsars to be $9\times 10^6$ years. After that, pulsars become dark neutron stars. According to the star formation history reconstructed from the chemical abundances in the Solar neighborhood \citep{Snaith2015}, the thick disk stage lasts approximately 4 –5 Gyr. It is then followed by a significant dip in star formation and a period of nearly constant star formation activity in the Milky Way thin disk during 8 Gyr.  Using this model, we find that the density of neutron stars in the solar neighborhood is about $1\times10^{-3} M_\odot/pc^3$. Our value of the volume density of neutron stars in the solar neighborhood is slightly higher than the value of the local volume density of neutron stars used by \citet{Kee15}. These authors based their estimate of the local volume density of neutron stars on the model of \citet{Sartore2010}, who assumed that the total number of neutron stars in the galaxy is equal to $10^9$. Considering that the Super Novae explosion is asymmetric, a neutron star receives a kick and likely leaves the area close to the midplane. Thus, we consider our estimation as an upper limit.

Black holes are a much less frequent product of stellar evolution. In general, to get a black hole as a final stage of evolution,  a star with a mass larger than that of a neutron star progenitor is required. The number of such stars decreases as a power-law function, while the average black hole mass can differ by a factor of 10 at most. Therefore, the density of black holes should be even smaller than the density of neutron stars.

\begin{deluxetable*}{l | c c}
\tablecaption{Local Mass Density}
\tablehead{
\colhead{Description} & \multicolumn{2}{c}{$\rho_{100}, M_\odot pc^{-3}$}
}
\startdata
$M > 0.2M_\odot$ & \multicolumn{2}{c}{$0.025_{-0.006}^{+0.012}$} \\
Binaries $M > 0.2M_\odot$ & \multicolumn{2}{c}{$0.0034\pm0.0003$} \\
$M < 0.2M_\odot$ &  $0.0089\pm0.0001$\tablenotemark{a} & $0.0058\pm0.0001$\tablenotemark{b} \\
White Dwarfs & \multicolumn{2}{c}{$0.00269\pm 0.00009$} \\
Neutron Stars and Black Holes & \multicolumn{2}{c}{$<0.001$} \\
\hline 
Total & $0.040_{-0.006}^{+0.012}$ & $0.037_{-0.006}^{+0.012}$\ 
\enddata
\tablenotetext{a}{with K01 IMF}
\tablenotetext{b}{with C05 IMF}
\end{deluxetable*}

\subsection{Midplane value}

Usually, the parameters of the disk model include the value of the mass density in the midplane. However, we wish to highlight that the choice of a function of the density profile can affect the final estimation of the mid-plane value. One should be careful when comparing the results based on the underlying model of the disk, because two widely used functions of the disk, such as exponential and $sech^2$, differ significantly in the mid-plane. Thus, we advise carefully constructing a fitting procedure because a vertical disk profile fitted at a large vertical distance might poorly represent the actual mid-plane mass density. We consider the $sech^2$ function to be more suitable following the theoretical modeling of an isothermal disk, as well as taking into account that the first derivative of the density distribution does not have a discontinuity at the mid-plane of the disk.

One should take into account the difference between the scale heights of the thin and thick disks, as well as the thick-to-thin-disk density ratio. The scale height of the thick disk is 3-4 times larger compared to that of the thin disk; thus, by changing the thick-to-thin-disk fraction, we change the total disk profile, making it steeper when the ratio is decreased. Another factor that should be taken into account is the position of the Sun relative to the midplane of the galactic disk. Depending on the vertical displacement value, the midplane value varies by a fraction of a percent.

In our study, we determine the value of the local mass density of stars which is free from model assumptions. 
Equation \ref{eq:rho} shows how one can compare the mass density in the miplane $\rho_0$ with a given vertical profile with our observational estimation $\rho_{100}$, assuming that the density does not change in the radial direction. Here, $z_0$ is the solar vertical offset. 

\begin{equation}
\label{eq:rho}
\begin{split}
\rho_{100} = & \frac{1}{\frac{4}{3}\pi \cdot100^3} \times \\
&\int_{0}^{100}2\pi RdR \int_{-\sqrt{100^2-R^2}-z_0}^{\sqrt{100^2-R^2}-z_0} \rho(z)dz
\end{split}
\end{equation}

We provide a midplane value of the mass density using recent estimates of the scale heights of the thin and thick disks, as well as the thick-to-thin-disk fraction from \citet{Viera2023}. The authors used the $sech^2$ function for both disk components, which fits well observational data up to 700 pc. Using their vertical disk profile with scale height of $279.76\pm12.49$ pc and $797.23\pm12.34$ pc with a thick-to-thin ratio of $0.750\pm0.049$ we get a midplane value of the mass density of $\rho_0 = 0.041_{-0.006}^{+0.012}M_\odot pc^{-3}$ for K01 IMF and $\rho_0 = 0.038_{-0.006}^{+0.012}M_\odot pc^{-3}$ for C05 IMF.

\citet{Everall2022} used a two-component exponential function to fit the disk. Their scale heights are $260\pm26$ pc for the thin disk and $693\pm121$ pc for the thick disk, with the thick-to-thin disk ratio of $0.141\pm0.075$. Adopting this function to our averaged value, we report that the midplane mass density is $\rho_{0}=0.045_{-0.007}^{+0.015} M_\odot pc^{-3}$ for K01 IMF and $\rho_{0}=0.042_{-0.007}^{+0.015} M_\odot pc^{-3}$ for C05 IMF. As can be seen, the midplane values here are higher than the ones derived using $sech^2$, even though they are normalized to the same average mass density $\rho_{100}$. This difference of around $10\%$ is simply due to the different behavior of the exponential and $sech^2$ functions around the mid-plane.

\subsection{Spatial density variations}

To explore the variations in density we select a subsample with absolute magnitudes ranging within $G_{abs} \in [7.5,15.5]$ inside the cylinder with a radius of $50\sqrt2$ pc and a height of $2\times50\sqrt2$ pc. The cylinder consists of 107224 stars, which is approximately $35\%$ of our sample.

First, we build a two-dimensional kernel density estimation (KDE) of the XY-plane using an Epanechnikov kernel with a smoothing length of 3 pc (see Fig.\ref{fig:spatial} left panel). To make variations clearly visible, we color-code the relative variations. We note that on such small scales, there are many substructures, the most noticeable overdensity located around $(-40,0)$ being the Hyades Open Cluster. We divide the XY-plane into four quadrants, and in each quadrant we compute the total number of stars and the total stellar mass (see Table \ref{table:var}). We find that there is a difference of around $3\%$ along the X coordinate in the total number and the total mass of stars. Considering an exponential radial profile with a scale length of 3 kpc, we expect the variation to be around $2\%$. We also find that the density variation along the Y coordinate is less than $1\%$.


Then we build a KDE of Z coordinate using an Epanechnikov kernel with a smoothing length of 10 pc (see Fig.\ref{fig:spatial} right panel). We plot thin disk profiles from \citet{Viera2023} and \citet{Everall2022}, however, considering that the range of Z coordinate of our sample is even smaller than a third of the scale height of the thin disk, we cannot conclude which profile is preferable, because even a smoothed KDE can be significantly affected by small-scale variations as indeed a peak at $z=-18.1$ related to the Haydes Open Cluster. The north-south difference in the total number and total mass of stars is around $2\%$.

\begin{figure*}
  \resizebox{\hsize}{!}{\includegraphics{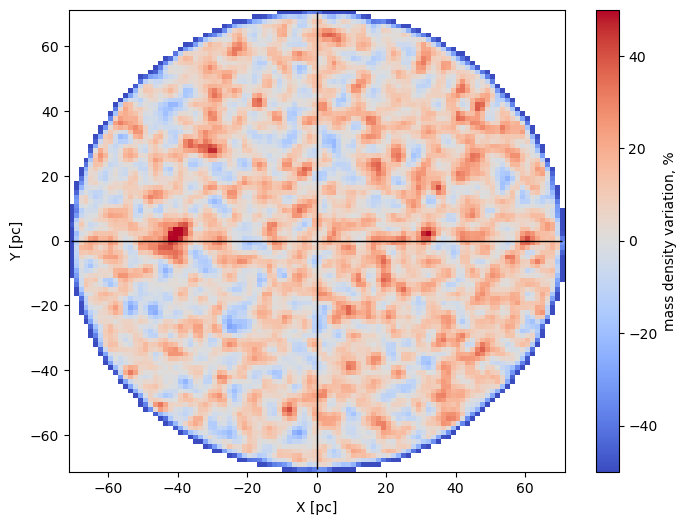}
    \includegraphics{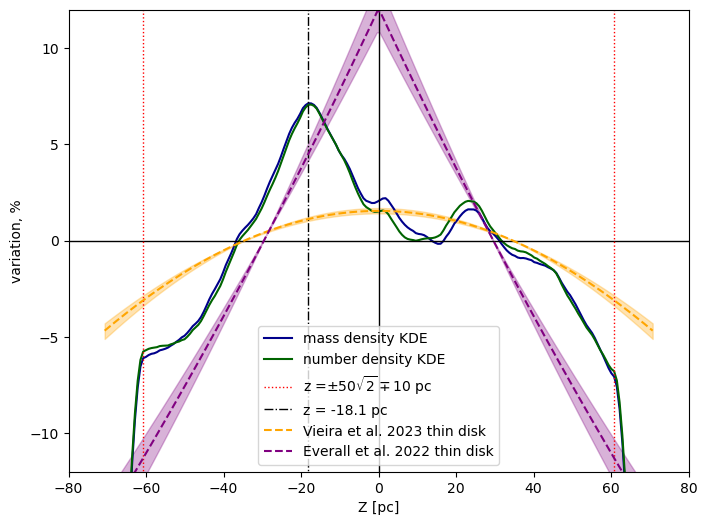}}
  \caption{Left panel: KDE with an Epanechnikov kernel and a smoothing length of 3 pc of mass density on the XY-plane color coded by mass density variations in percents. Right panel: KDE with an Epanechnikov kernel and a smoothing length of 10 pc of mass density in blue, number density in green. The red dotted lines represent the border after which KDE drops because of smoothing. The black dotted-dashed line shows the Z coordinate of the maximum value of KDE. The dashed orange and purple lines with corresponding filled area are the vertical profiles of the thin disk from \citet{Viera2023} and \citet{Everall2022} with their errors.}
  \label{fig:spatial}
\end{figure*}

\begin{deluxetable*}{l | c c}
\label{table:var}
\tablecaption{Spatial variation}
\tablehead{
\colhead{Description} & Total number & Total mass, $M_\odot$ 
}
\startdata
$(X > 0)\&(Y >0)$ &  $27236$ & $7183\pm1090$\\
$(X < 0)\&(Y >0)$ & $26614$ & $7075\pm1065$ \\
$(X < 0)\&(Y <0)$ & $26123$ & $6919\pm 1046$ \\
$(X > 0)\&(Y <0)$ & $27251$ & $7201\pm1091$ \\
\hline
$(Z > 0)$ & $53200$ & $14066\pm2130$ \\
$(Z < 0)$ & $54024$ & $14313\pm2162$ \\
\enddata
\end{deluxetable*}
 
\section{Discussion} \label{sec:disc}\

In the pre-Gaia era several attempts have been made to estimate the density of baryonic matter in the solar neighborhood. \citet{Flynn} using Hipparcos and Tycho surveys combined with the CNS4 catalog estimated the baryonic mass density in the solar neighborhood of $\rho_{0}=0.0415M_\odot pc^{-3}$. Later, \citet{Kee15} reported a slightly higher value of $\rho_{0}=0.043M_\odot pc^{-3}$, explaining the difference from the alleged study by \citet{Flynn} due to the use of more complete updated data. Both works present a comprehensive analysis of the available data at that time.

\citet{McMillan_2011}} used the Bayesian fitting to construct the Galactic model. Taking into account kinematical data, they reported the mass density of $\rho_{0}=0.087\pm0.010M_\odot pc^{-3}$. In later works \citep{McMillan_2016,McMillan2017} slightly different best-fit parameters were reported that result in a local midplane density of around $0.052M_\odot pc^{-3}$, which is consistent with our upper limit. The double exponential disk profile was assumed, which could explain the higher value compared to previous estimates.

\citet{Bovy17} using Gaia DR1 data constructed the vertical density profiles of the disk for different stellar types, concluding that the $sech^2$ function fits the observational data better than the exponential density profile and found the midplane stellar density $\rho_{0}=0.040\pm0.002M_\odot pc^{-3}$, which is close to our estimate.

\citet{Everall2022} based their estimate of the local stellar density on the Gaia EDR3 data. The authors fitted the model parameter to the observational data, assuming that the disk has a double exponential profile taking into account the solar vertical offset, dust extinction, and parallax offset (\citet{Everall2022} their Sect. 5), concluded that systematic errors are much higher compared to statistical ones and reported the local stellar mass density of $\rho_{0}=0.0366\pm0.0052 M_\odot pc^{-3}$. Excluding the contribution of the WD population from our mass density analysis we find that the midplane density of stars in model with double exponential disk profile with scale heights of  $260\pm26$ pc  and $693\pm121$ pc is $\rho_{0}=0.042_{-0.007}^{+0.014} M\odot pc^{-3}$ for K01 IMF and $\rho_{0}=0.038_{-0.007}^{+0.014} M\odot pc^{-3}$ for C05 IMF being consistent with result of \citet{Everall2022} within the errors.

In general, our results are in agreement with previous estimates. However, we note that our error bars are larger because we do not assume a disk profile. We provide a realistic estimation of errors in mass derived using empirical relations. We find that the mass density of the WD stars is lower compared to previous estimates. \citet{Kee15}, for instance, reported that the density of WD stars is $0.0056 M_\odot pc^{-3}$, which is twice as high as our estimate. We note here that the estimate of \citet{Kee15} was obtained using the fraction of WD to M dwarfs, derived from stellar tracks of assumed age and IMF, while our estimate is based on observational data.

\begin{figure}
  \resizebox{\hsize}{!}{\includegraphics{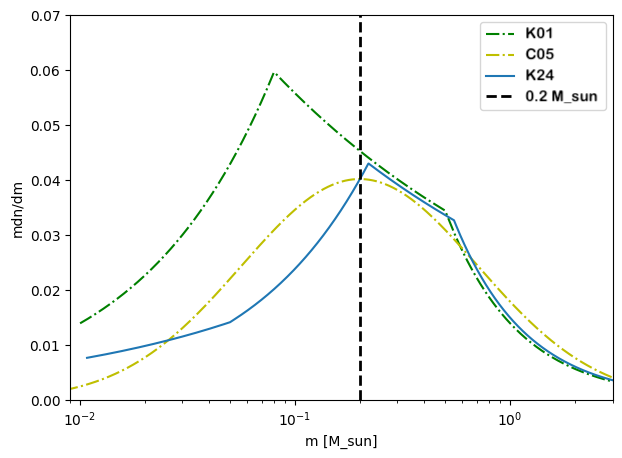}}
  \caption{Comparison of the mass contribution between IMFs. The green line is K01 IMF. The yellow line is C05 IMF, while the blue line is K24 IMF. The black vertical line shows the limit of $0.2M_\odot$.}
  \label{fig:mass_hist_kirkpatrick}
\end{figure}

Recently, K24 used Gaia DR3 data and measured the number density of individual stars within the Sun-centered sphere of 20 pc radius. K24 took into account the mass of binary and multiple stellar systems by decomposing the binaries or higher-order systems into separate components. As a result, they got about 3600 individual stars that were used to construct the distribution of stars in the close vicinity of the Sun as a function of their masses. K24 gave the empirical expression that describes the density of stars per $pc^3$ per $M_\odot$ for different mass intervals of individual stars.  Using this expression, we plotted in Fig. \ref{fig:mass_hist_kirkpatrick} the dependence of the mass density of the stellar component as a function of the mass of individual stars for the K24 mass function. For comparison, we also plot similar distributions for mass functions determined in our paper using K01 IMF and C05 IMF. As one can see, all distributions are close to each other for stellar masses larger than $0.2 M_\odot$. For the low-mass end of the mass function, the distributions differ considerably. 
We find that the mass density of stars within the mass interval $[0.0001; 0.2] M_\odot$ is $0.0046 M_\odot pc^{-3}$ for K24, which is comparable to our estimate using C05 IMF $0.0058 M_\odot pc^{-3}$. However, the value of K24 is considerably lower compared to the same value for the K01 IMF, which gives the value of $0.0089 M_\odot pc^{-3}$ for stars within the same interval of stellar masses.


Following recent works on the spiral structure of the Milky Way, we should mention that at a scale of several hundreds of parsecs, there is a variation of density related to the spiral arms seen by O-B stars \citep{Poggio2021,Drimmel2023,Ge_2024}. \citet{Poggio2021} suggest that the Sun is in the region of lower density, while \citet{Drimmel2023} and \citet{Ge_2024} showed that the position of the Sun corresponds to the mean or slightly overdense region. Such a variation can, in principle, affect the applicability of our estimate of the local stellar mass density for large-scale dynamical modeling, however, we should stress here that the O-B star population is a small subset, and understanding the density variation in other stellar populations requires deeper investigations.

\section{Summary}
\label{sec:sum}

We determined the mass density of stars and stellar remnants in the solar neighborhood within a 100 pc sphere using the Gaia Catalog of Nearby Stars based \citep{Smart_GCNS_2021} on Gaia EDR3 together with the catalog of White Dwarf stars from \citet{O’Brien_WD40_10.1093/mnras/stad3773} and a recent estimate of surface density of neutron stars from \citet{Xie_2024_NS}. To compute masses of individual stars in the sample, we used the mass-magnitude or mass-stellar parameter relations. The mass function was corrected for incompleteness and unresolved close binary stars. To correct for the low-mass part of the stellar mass function, we fit the observed mass function to the IMF of \citet{Kroupa2001MNRAS.322..231K} and the IMF of \citet{chabrier2005initial}. We find that using the Kroupa IMF, the total stellar density in the solar neighborhood is $\rho_{100} = 0.040^{+0.012}_{ -0.006}M_\odot pc^{-3}$, while with the Chabrier IMF the value is $\rho_{100} = 0.037^{+0.012}_{-0.006}M_\odot pc^{-3}$. 

We stress that our estimation is independent of a Galactic model, and therefore can be used with any model deriving midplane mass density $\rho_0$ from our averaged local mass density $\rho_{100}$. We present here the example of such normalization, deriving midplane densities with the vertical profiles of \citet{Viera2023} and \citet{Everall2022}. In the first case, we obtain $\rho_{0}=0.041_{-0.006}^{+0.012} M_\odot pc^{-3}$ for K01 IMF and $\rho_{0}=0.038_{-0.006}^{+0.012} M_\odot pc^{-3}$ for C05 IMF, in the second case, $\rho_{0}=0.045_{-0.007}^{+0.014} M_\odot pc^{-3}$ for K01 IMF and $\rho_{0}=0.042_{-0.007}^{+0.014} M_\odot pc^{-3}$ for C05 IMF.

We explored density variations within the complete subset. However, we report that the volume of our study is too small to draw meaningful conclusions about the density profile.





\begin{acknowledgments}
The authors thank Antonella Vallenari for useful comments and suggestions. The authors are grateful to the referee for improving the article.
VK and RT thank Ministry of Science and Higher Education of Russian Federation for support under state contract GZ0110/23-10-IF.
This work presents results from the European Space Agency (ESA) space
mission Gaia. Gaia data are being processed by the Gaia Data Processing and Analysis Consortium(DPAC). Funding for the DPAC is provided by national institutions, in particular the institutions participating in the Gaia Multi-Lateral Agreement (MLA). The Gaia mission website is https://www.cosmos.esa.int/gaia, accessed on 25 April 2025. The Gaia archive website is https://archives.esac.esa.int/gaia, accessed on 25 April 2025.
\end{acknowledgments}

%

\vspace{5mm}


\software{GaiaUnlimited \citep{Castro-Ginard_ruwe},  
          TOPCAT \citep{TOPCAT}, 
          Scipy \citep{2020SciPy-NMeth}
          }



\appendix

\section{Initial cut} \label{sec:initial_cut}

\begin{figure*}
  \resizebox{\hsize}{!}{\includegraphics{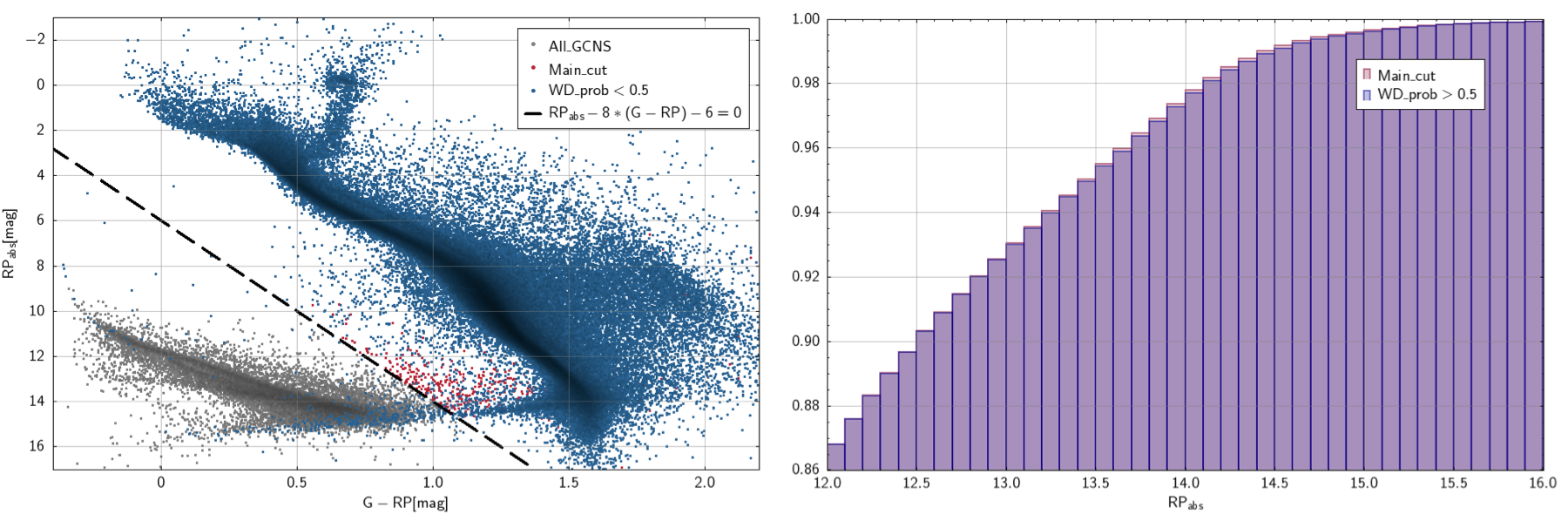}}
  \caption{Left panel: color absolute magnitude diagram for the absolute magnitude $RP_{abs}$ and color index $G-RP$. All stars from the GCNS are plotted in grey, red points are the sample used in this article with the cut plotted in black, while blue points are the sample using $WD\_prob<0.5$ to remove the WD population. Right panel: normalized cumulative histogram of absolute magnitude RP for the samples as before.}
  \label{fig:init_cut}
\end{figure*}

One might be cautious about the cut used to remove the WD population. Here, the value of $WD\_prob$ was used instead to remove WD stars from the GCNS catalog. We identified WD stars with $WD\_prob \geq 0.5$, and therefore a subsample of the GCNS without WD stars with $WD\_prob < 0.5$. The resulting total number of stars is 302457 (instead of 302449). The main difference is for the absolute magnitude $RP_{abs}>12$ (see Fig. \ref{fig:init_cut} left panel), however as one can notice from the cumulative histogram, the difference is negligible (see Fig. \ref{fig:init_cut} right panel).


\section{Individual masses} \label{sec:ind_mass}

Here we provide four examples of masses estimated as described in Sect.\ref{sec:mass}. 

\begin{figure*}
  \resizebox{\hsize}{!}{\includegraphics{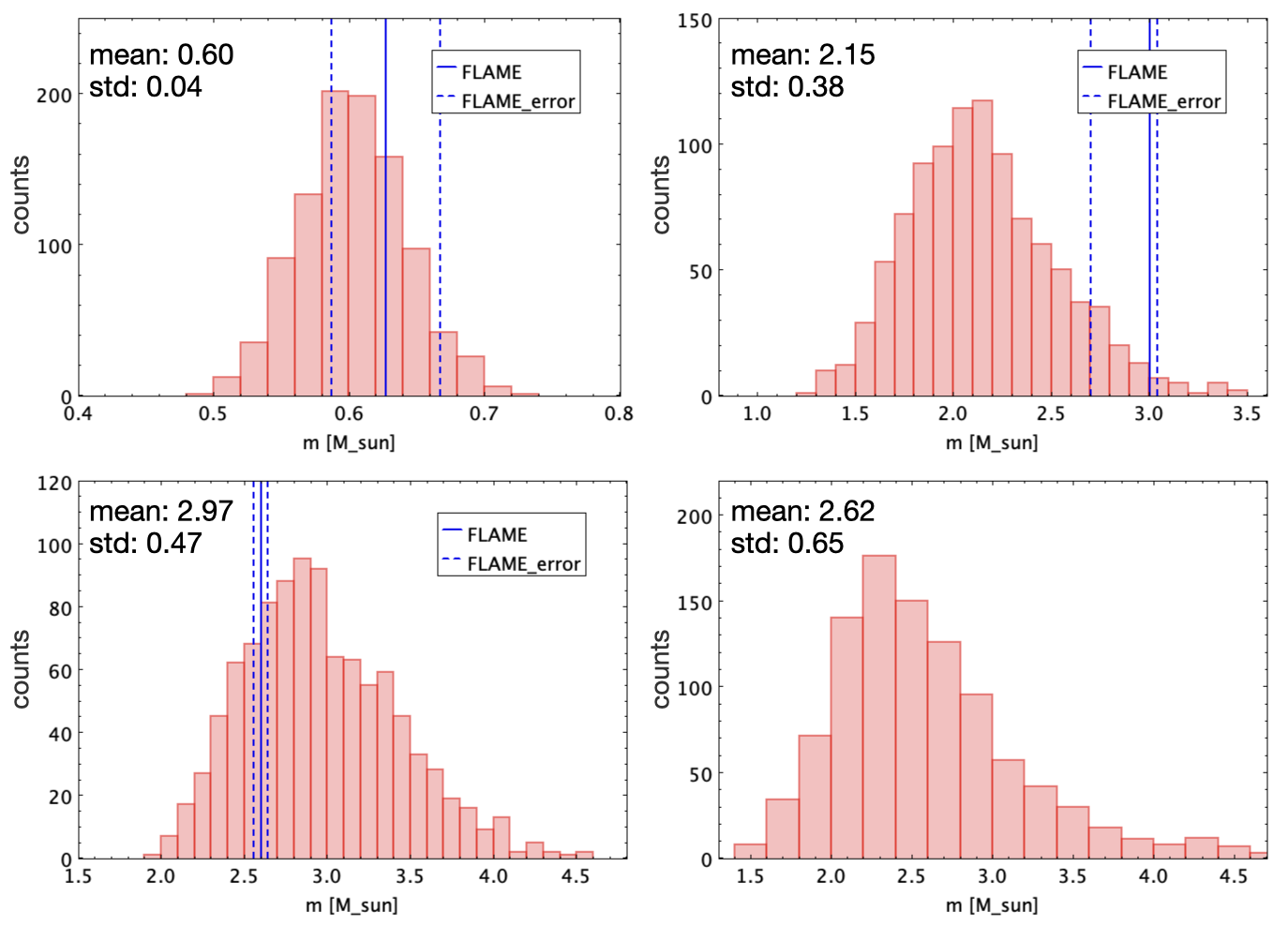}}
  \caption{Histograms of masses for 4 stars with computed $mean$ and $std$ of the distribution. The solid and dashed blue vertical lines are reported mean and errors of FLAME masses respectively.}
  \label{fig:ind_mass}
\end{figure*}

The first star ($source\_id = 6343505725112760320$) is a bad \textit{ruwe} ($ruwe=7.41$) MS star with available FLAME mass estimation (see Fig. \ref{fig:ind_mass} upper left panel). For this star, the G22 relation was used to obtain the mass. As one can see, the distribution of mass is close to a Gaussian and the difference between the mean value and the FLAME mass is in one standard deviation.

The second star ($source\_id = 4423234230853333632$) is a Red Clump star with available FLAME mass estimation (see Fig. \ref{fig:ind_mass} upper right panel). The mass of this star was derived from the M18 relation. For this star, we can see non-Gaussian tails towards larger masses. Also, one can see that the $std$ is much larger with respect to the MS star. The FLAME mass differs significantly here; however, it is mentioned that evolved stars with masses more than $2M_\odot$ are probably misclassified and their masses should not be used ($flags\_flame=10$).  

The next star ($source\_id = 1337664063942689792$) is a star in the upper MS, again, with available FLAME mass (see Fig. \ref{fig:ind_mass} lower left panel). The mass was derived from the M18 relation. As in the previous example, one can see a tail towards larger masses, and a large $std$. The FLAME mass ($flags\_flame=0$) lies in one $std$ deviation from the mean.

The final example ($source\_id =1274795260380782080$) is a star from the upper MS, but this time without stellar parameters available and consequently without the FLAME mass (see Fig. \ref{fig:ind_mass} lower right panel). For this star, the mass was obtained as an average mass of the cell. Such stars present the largest errors in the estimated mass; however, as mentioned before, the total sample consists of a few percent of such sources.


\section{The Coma Berenices Open Cluster} \label{sec:Coma_Ber}

\begin{figure*}
  \resizebox{\hsize}{!}{\includegraphics{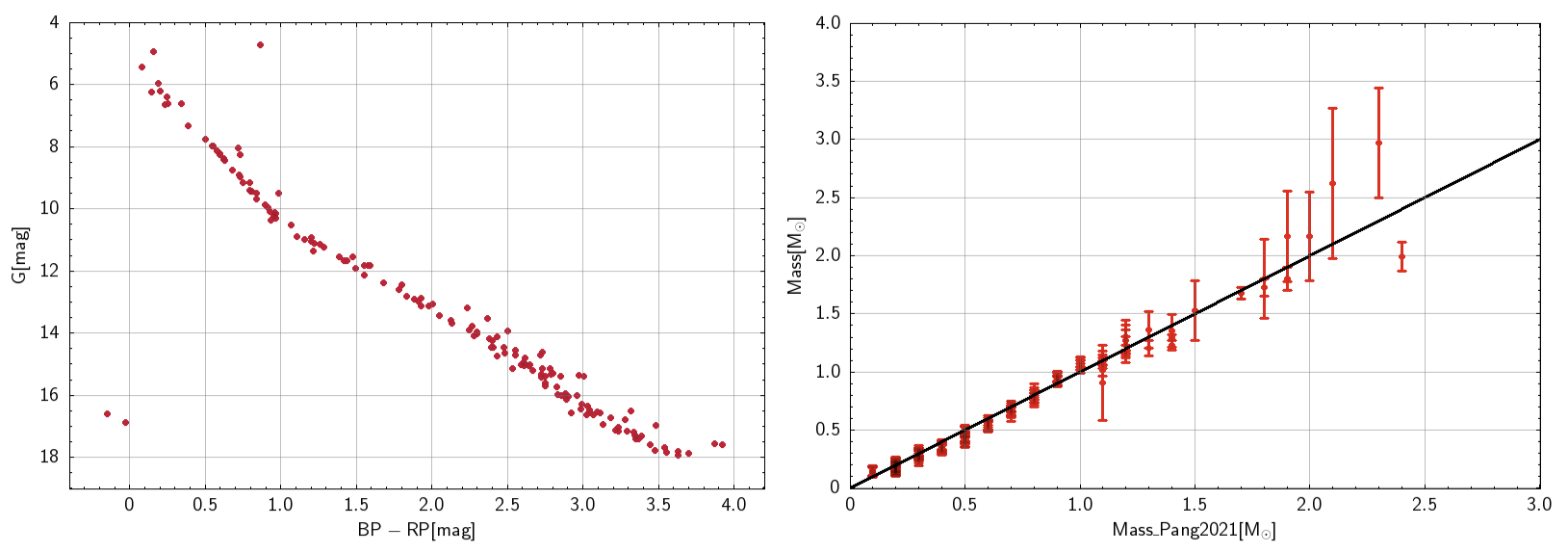}}
  \caption{Left panel: Color magnitude diagram for $G$ magnitude and $BP-RP$ color index of the Coma Berenices OC. Right panel: comparison between individual masses derived in this work to the masses from \citet{Pang_2021}. The black line is 1:1 correspondence.}
  \label{fig:Coma_ber}
\end{figure*}

Here we show how the masses derived by our method correspond to the masses derived from isochrone fitting. We use members of the Coma Berenices Open Cluster (OC) from \citet{Pang_2021}. 
We cross-matched their members with our sample. We found 156 matched stars out of the 158 reported by \citet{Pang_2021}. The missing two stars are WD stars, as one can clearly see from the left panel of Fig. \ref{fig:Coma_ber}. 
\citet{Pang_2021} used PARSEC isochrones to fit the OC color-magnitude diagram. The masses were derived after the best-fit isochrone was found. As one can see from the right panel of the Fig. \ref{fig:Coma_ber}, the masses derived from our method and the masses derived from the isochrone are in good agreement. The total mass of the Coma Berenices OC members reported by \citet{Pang_2021} is $101.6 M_{\odot}$, while the total mass derived from our approach, considering two missing WD stars with mass $0.6 M_{\odot}$, is $99.6 \pm 9.5 M_{\odot}$.


\bibliography{sample631}{}
\bibliographystyle{aasjournal}



\end{document}